\documentclass[apjl,twocolumn,linenumbers]{emulateapj_mod}
\usepackage{epsfig,apjfonts,mathptmx,amssymb}

\def\gtsima{$\; \buildrel > \over \sim \;$}
\def\ltsima{$\; \buildrel < \over \sim \;$}
\def\prosima{$\; \buildrel \propto \over \sim \;$}
\def\gsim{\lower.5ex\hbox{\gtsima}}
\def\lsim{\lower.5ex\hbox{\ltsima}}
\def\simgt{\lower.5ex\hbox{\gtsima}}
\def\simlt{\lower.5ex\hbox{\ltsima}}
\def\simpr{\lower.5ex\hbox{\prosima}}

\def\h1{$h^{-1}$}
\def\eeq{\end{equation}}
\def\beq{\begin{equation}}
\def\24mu{24\,$\mu{\rm m}$}
\def\70mu{70\,$\mu{\rm m}$}
\def\8mu{8\,$\mu{\rm m}$}

\def\lya{Ly$\alpha$}

\submitted{ApJ Letters, in press}
\shorttitle{Cold-stream to hot-accretion transition at $2<z<3.3$}
\shortauthors{Daddi et al.}
\begin{document}

\title{Evidence for the cold-stream to hot-accretion transition\\ as traced by \lya\ emission from groups and clusters at 2$<$\lowercase{z}$<$3.3}

\author{
E. Daddi\altaffilmark{1}, 
R.~M. Rich\altaffilmark{2},
F. Valentino\altaffilmark{3,4},
S. Jin\altaffilmark{3,5},
I. Delvecchio\altaffilmark{6},
D. Liu\altaffilmark{7},
V. Strazzullo\altaffilmark{8},
J. Neill\altaffilmark{9},
R. Gobat\altaffilmark{10},\\
A. Finoguenov\altaffilmark{11},
F. Bournaud\altaffilmark{1},
D. Elbaz\altaffilmark{1},
B.~S. Kalita\altaffilmark{1},
D. O'Sullivan\altaffilmark{9},
T. Wang\altaffilmark{12}
}

\altaffiltext{1}{CEA, IRFU, DAp, AIM, Universit\'e Paris-Saclay, Universit\'e de Paris,  Sorbonne Paris Cit\'e, CNRS, F-91191 Gif-sur-Yvette, France}
\altaffiltext{2}{Department of Physics \& Astronomy, University of California Los Angeles, 430 Portola Plaza, Los Angeles, CA 90095, USA}
\altaffiltext{3}{Cosmic Dawn Center (DAWN) }
\altaffiltext{4}{Niels Bohr Institute, University of Copenhagen, Jagtvej 128, DK-2200, Copenhagen N, Denmark}
\altaffiltext{5}{DTU-Space, Technical University of Denmark, Elektrovej 327, DK-2800 Kgs. Lyngby, Denmark}
\altaffiltext{6}{INAF-Osservatorio Astronomico di Brera, via Brera 28, 20121, Milano, Italy}
\altaffiltext{7}{Max-Planck-Institut fur extraterrestrische Physik (MPE), Giessenbachstr. 1, 85748 Garching, Germany}
\altaffiltext{8}{Dipartimento di Fisica, Universit\'a di Trieste, Via Tiepolo 11, I-34143 Trieste, Italy}
\altaffiltext{9}{California Institute of Technology, 1216 East California Boulevard, Pasadena, California 91125, USA.}
\altaffiltext{10}{Instituto de Fisica, Pontificia Universidad Catolica de Valparaiso, Casilla, 4059, Valparaiso, Chile}
\altaffiltext{11}{Department of Physics, University of Helsinki, Gustaf H{\"a}llstr{\"o}min katu 2, FI-00014 Helsinki, Finland}
\altaffiltext{12}{School of Astronomy and Space Science, Nanjing University, Nanjing 210093, People's Republic of China}

\begin{abstract}
We present  Keck Cosmic Web Imager  (KCWI) observations of giant \lya\ halos  surrounding 9 galaxy groups and clusters at $2<z<3.3$, including { five new detections and one upper limit}. We find observational evidence for the { cold-stream} to hot-accretion  transition predicted by theory by  measuring a decrease in the ratio between the spatially extended \lya\ luminosity and the expected baryonic accretion rate (BAR), with increasing elongation above the transition mass ($M_{\rm stream}$). This implies a modulation of the share of BAR that remains cold  diminishing quasi-linearly (logarithmic slope of $0.97\pm0.19$, $5\sigma$ significance) with the   halo to $M_{\rm stream}$ mass ratio.  The integrated star-formation rates (SFRs) and AGN bolometric luminosities  display a { potentially} consistent decrease, albeit significant only at $2.6\sigma$ and $1.3\sigma$, respectively. The higher scatter in these tracers suggests the \lya\ emission might be mostly a direct product of cold accretion in these structures rather than indirect, mediated by outflows and photo-ionization from SFR and AGNs; this is also supported by energetics considerations.
{ Below $M_{\rm stream}$ (cold-stream regime)} we measure $L_{\rm Ly\alpha}/\mbox{BAR}=10^{40.51\pm0.16}$~erg~s$^{-1}$~$M_\odot^{-1}$~yr, consistent with  predictions, and SFR/BAR$=10^{-0.54\pm0.23}$: on average $30_{-10}^{+20}$\% of the cold streams go into stars. { Above $M_{\rm stream}$ (hot-accretion regime),}  $L_{\rm Ly\alpha}$ is set by $M_{\rm stream}$ (within 0.2~dex scatter in our sample), independent of the halo mass but rising tenfold from $z=2$~to~3. 
\end{abstract}


\section{Introduction}

{ It has long been understood from theory that galaxies in dark matter halos below $M_{\rm shock}\approx10^{12}M_\odot$ are fed by cold accretion, delivering gas ready to form stars (White \& Frenk 1991; Birnboim \& Dekel 2003; Keres et al. 2005; Dekel \& Birnboim 2006; DB06 hereafter),  driving high SFRs in distant galaxies (e.g. Genel et al 2008). 
Above $M_{\rm shock}$, cooling times are longer than dynamical times and shocks can efficiently heat incoming baryons. 

However, numerical simulations and analytical work from Dekel \& Birnboim (2006) first showed that cold accretion continues to penetrate at high redshifts in the form of cold streams even above $M_{\rm shock}$, in massive halos located at the intersection of multiple  and dense filaments, narrower than the halos they accrete onto. This is crucial for feeding even more massive galaxies (e.g., Daddi et al. 2007) residing in massive halos at high redshifts (e.g., Bethermin et al. 2014).}
Subsequent numerical and analytical modelling developed this theory  with respect to stream  
stability (Nelson et al. 2015; Mandelker et al. 2019), multi-phase properties (Cornuault et al. 2018), additional inner-halo cooling (Mandelker et al. 2019, 2020; Zinger et al. 2018), angular momentum transfer to galaxies (Danovich et al. 2015). 
{ This {\em cold stream} mode should be effective upto an evolving halo mass  $M_{\rm stream}(z)$, expected of order of $10^{12.5}M_\odot$ at $z=2$ and growing to $10^{13.5}M_\odot$ at $z=3$ (DB06), and rapidly diluted and disappearing at even higher masses. 

Observational confirmation is still lacking for cold streams, and for evidence that the $M_{\rm stream}$ transition affects gas accretion observables.}
Cold streams are predicted to be best detectable via their collision powered \lya\ emission (Dijkstra et al 2009; Goerdt et al. 2010; Rosdhal \& Blaizot 2012). However, deep \lya\ observations of distant massive groups and clusters are still scarce.
In this letter we present results from our ongoing KCWI survey targeting 9 massive galaxy structures at $2<z<3.5$, providing \lya-based  evidence of the predicted dilution of cold streams across $M_{\rm stream}$.
We adopt concordance cosmology (0.3; 0.7; 70) and a Chabrier IMF.

\begin{figure*}[ht]
\centering
\vspace{-0.3truecm}

\includegraphics[width=6.3cm,angle=90]{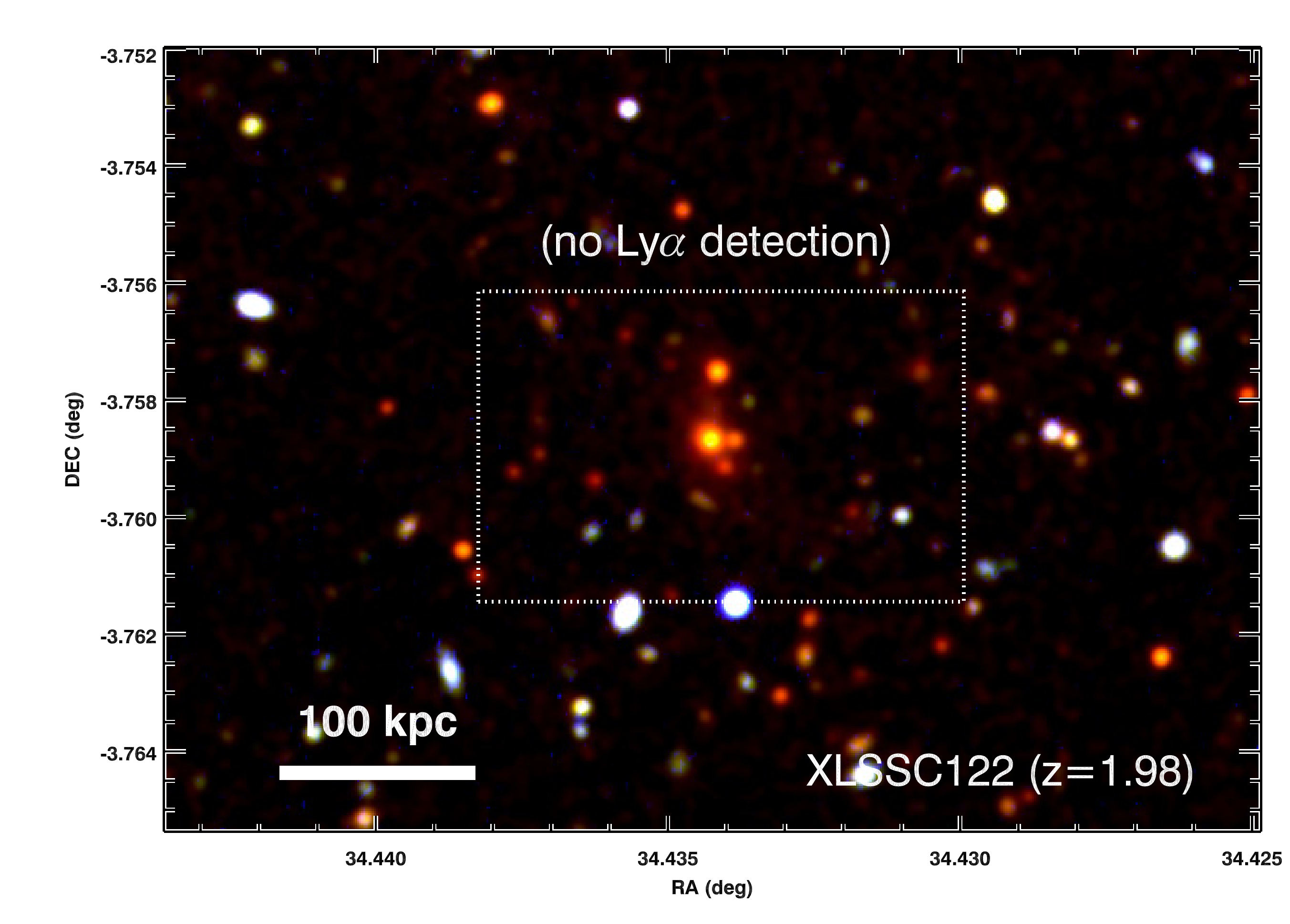}
\hspace{-0.6truecm} 
\includegraphics[width=6.3cm,angle=90]{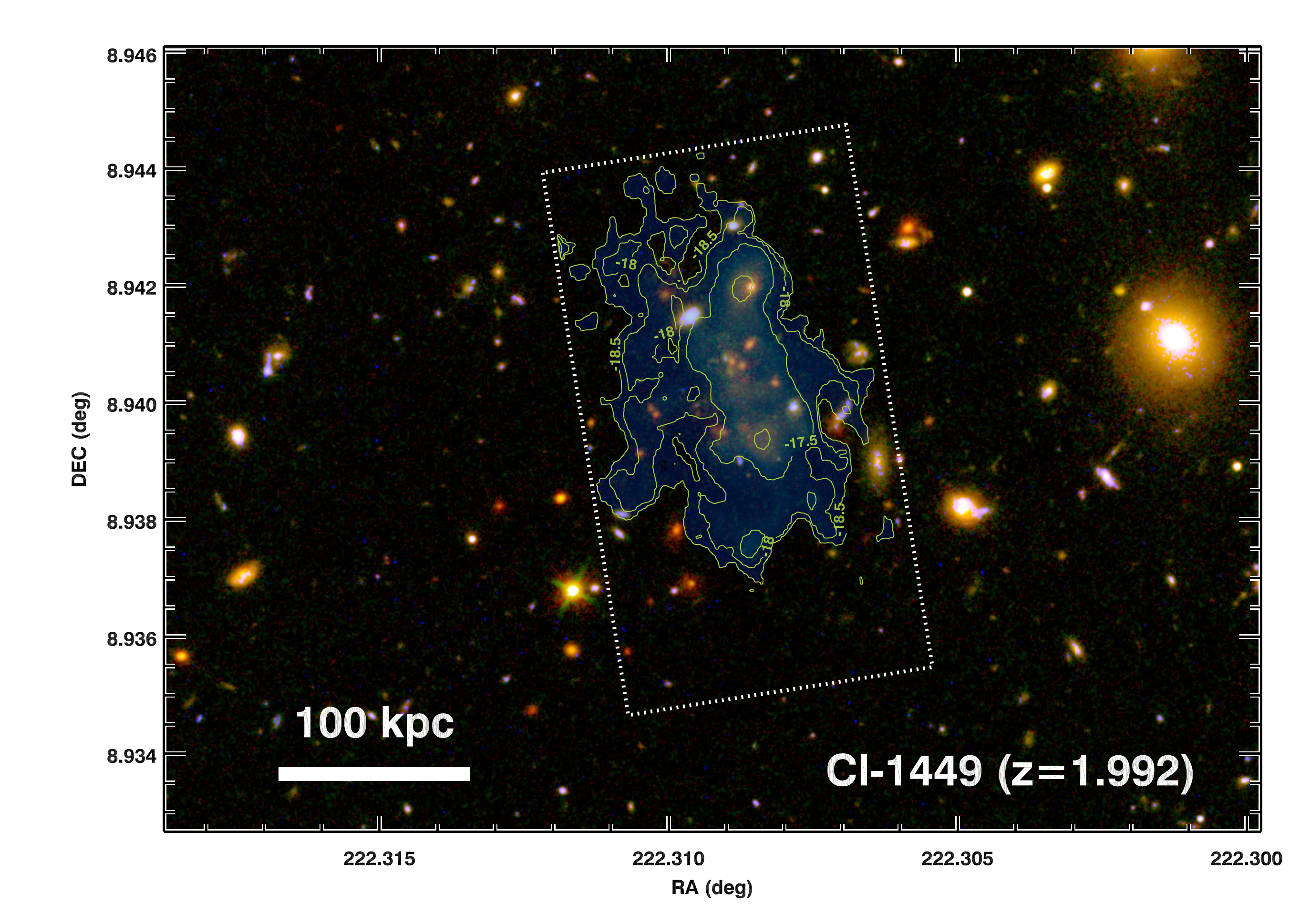}
\vspace{-0.3truecm}

\includegraphics[width=6.3cm,angle=90]{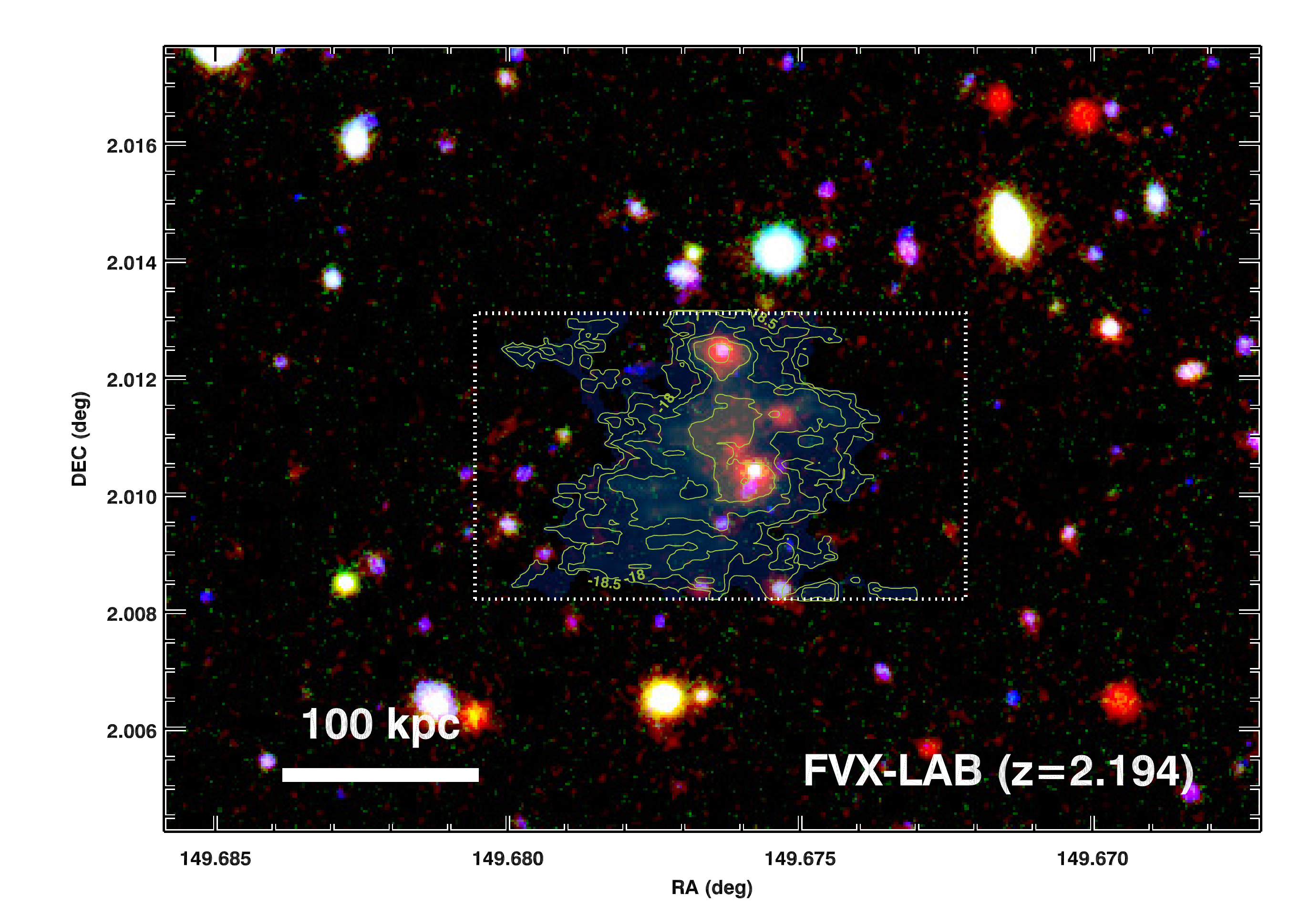}
\hspace{-0.6truecm}
\includegraphics[width=6.3cm,angle=90]{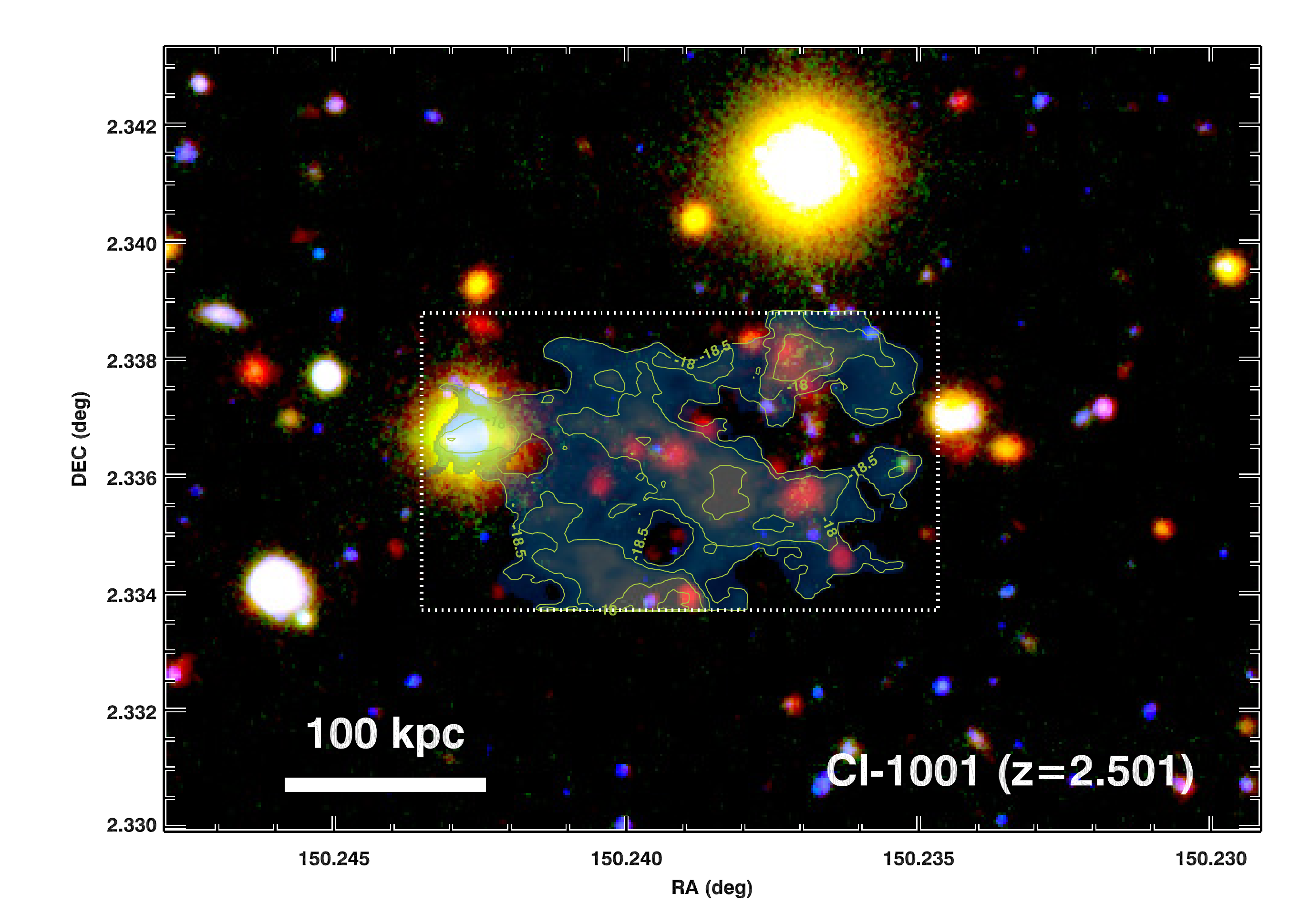}
\vspace{-0.3truecm}

\includegraphics[width=6.3cm,angle=90]{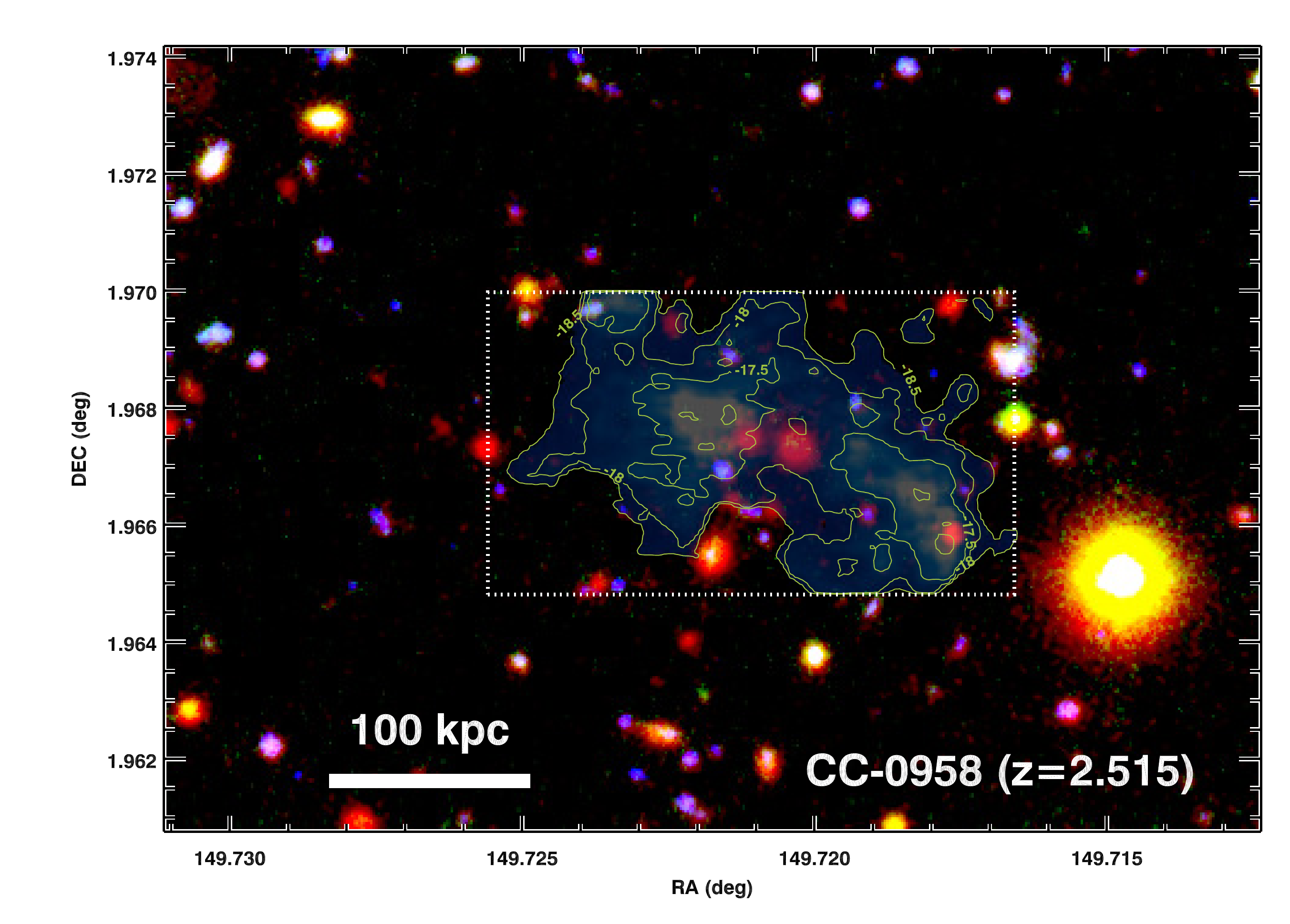}
\hspace{-0.6truecm}
\includegraphics[width=6.3cm,angle=90]{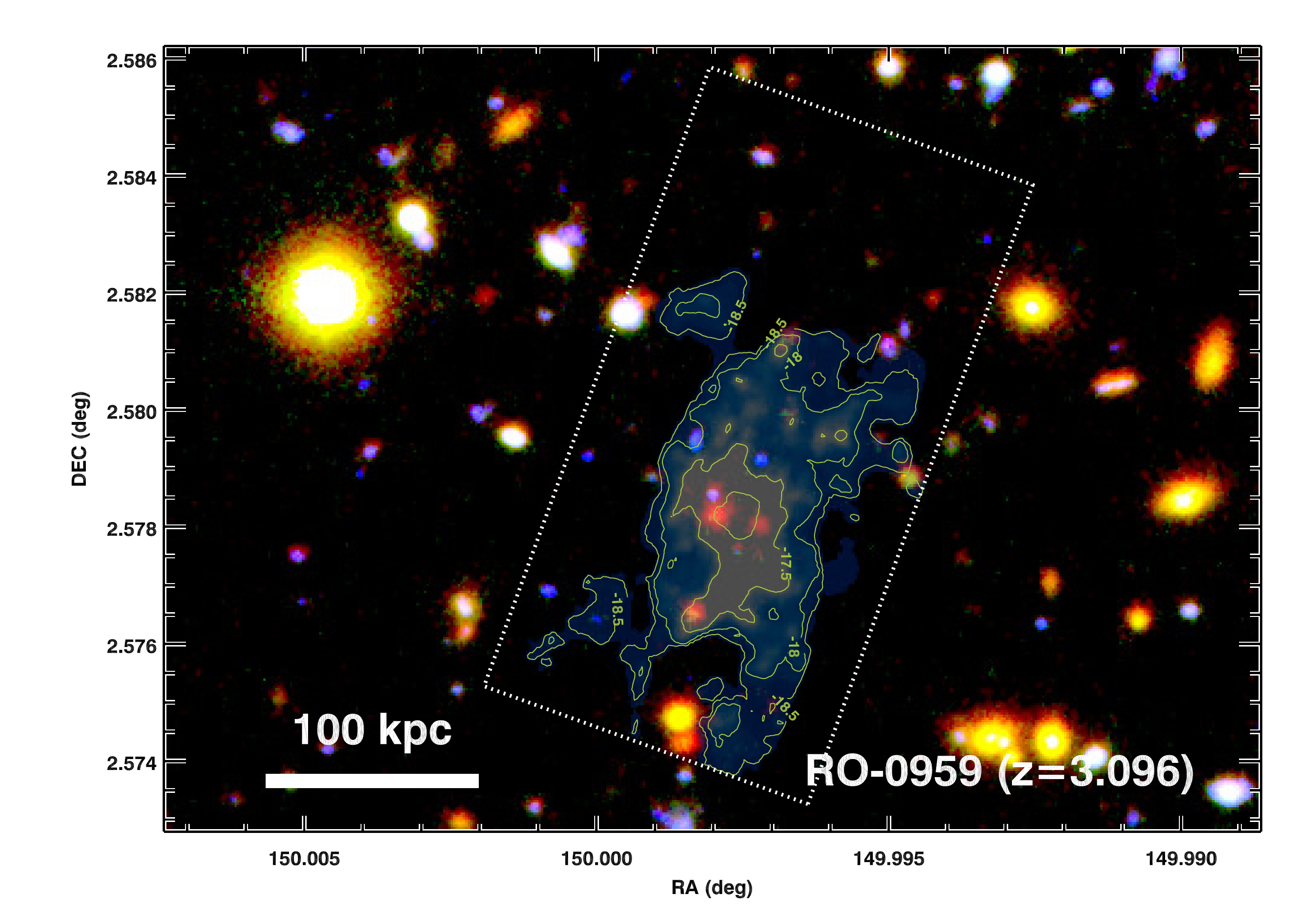}
\vspace{-0.3truecm}

\includegraphics[width=6.3cm,angle=90]{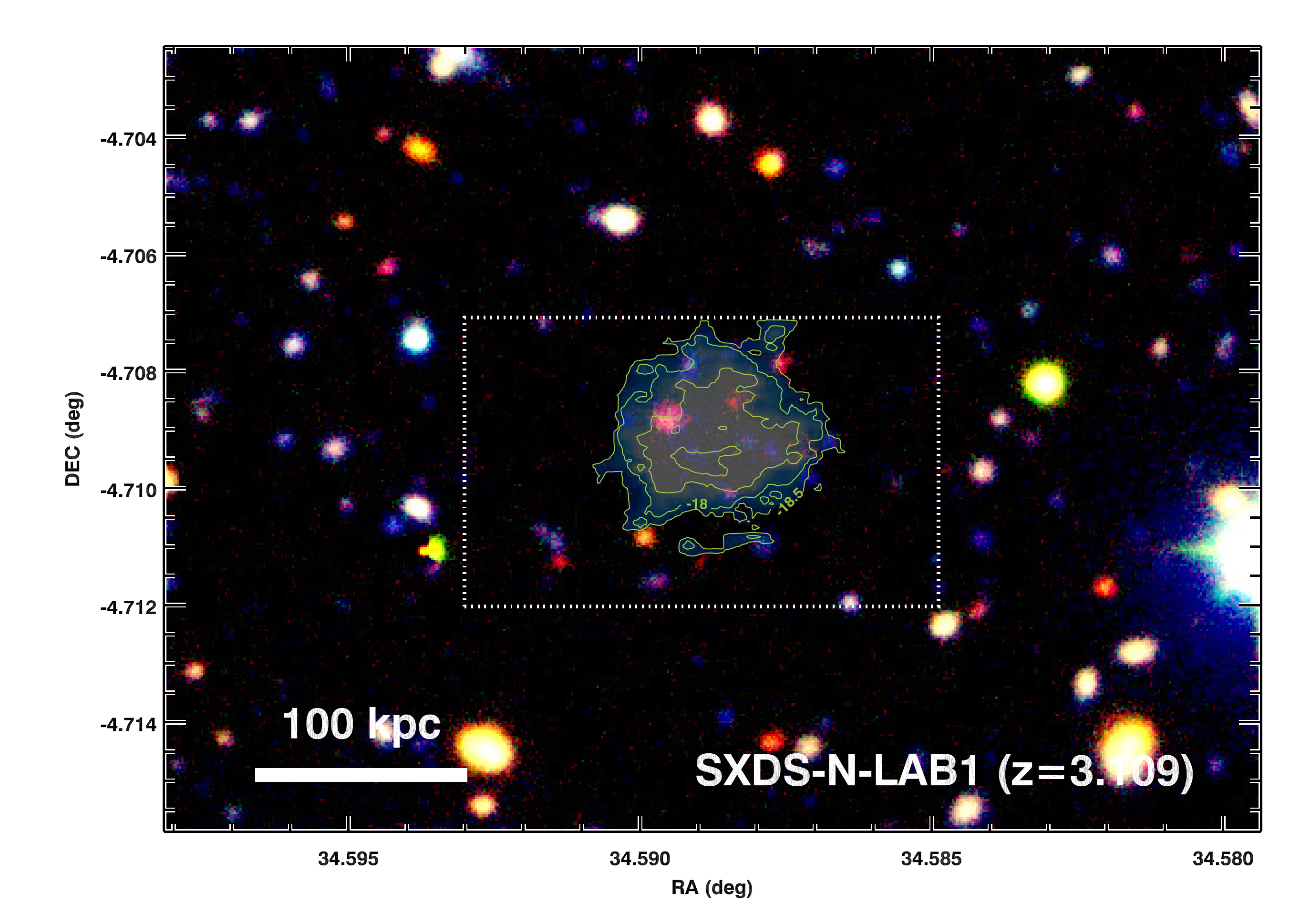}
\hspace{-0.6truecm}
\includegraphics[width=6.3cm,angle=90]{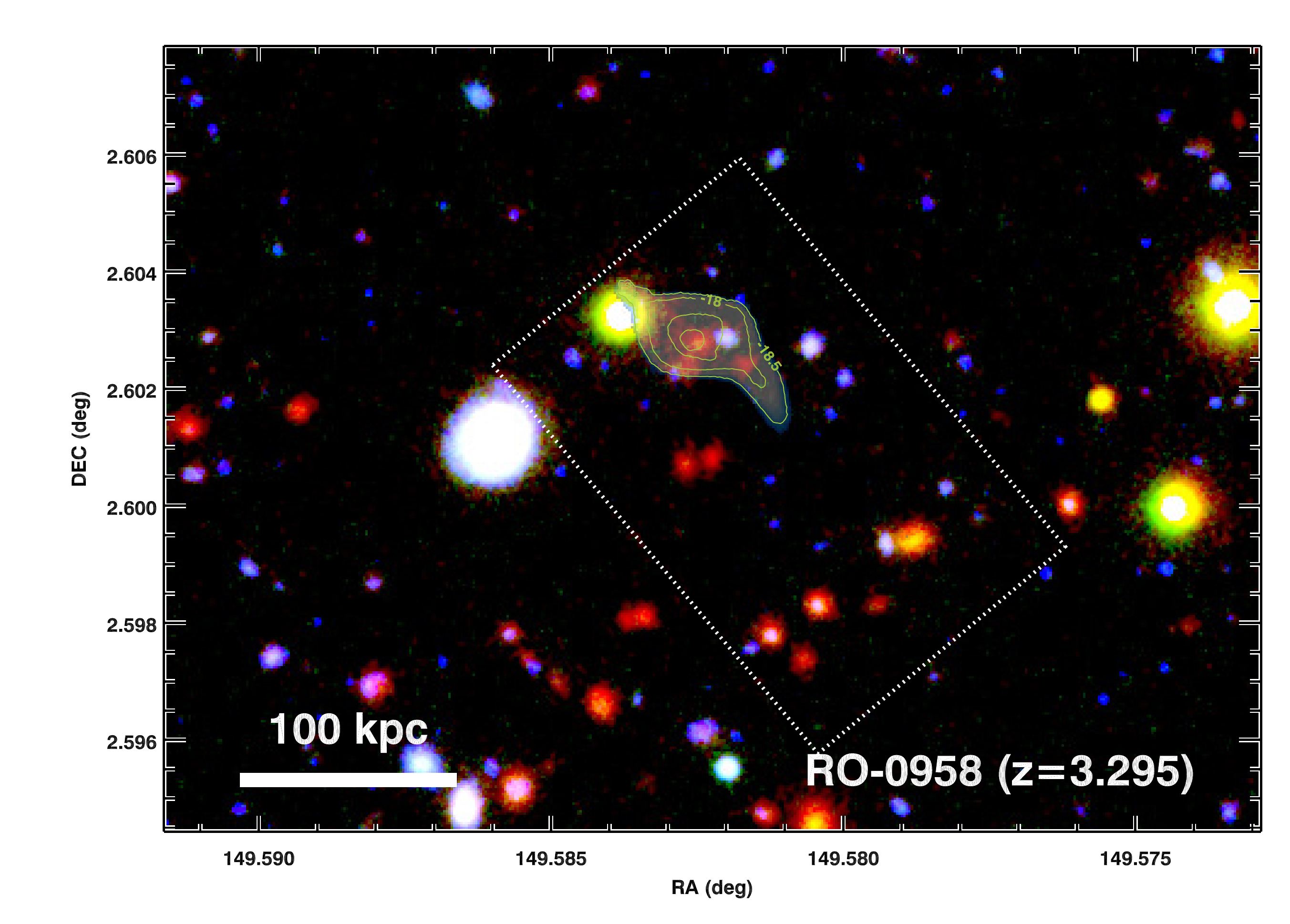}
\vspace{-0.3truecm}

\caption{Color images of spectroscopically confirmed targets for which KCWI \lya\ observations are first presented here (BzK for most, or close variations; North is up and East is left). Cl-1449 is from HST, the rest is ground-based. See D21 for a similar RO-1001 image.  The blue soft layer shows \lya\ emission, with contours displayed in log steps from -18.5 to -17.5  erg~s$^{-1}$~arcsec$^{-2}$ as labeled. The dotted lines show the KCWI field. The orientation of the RO-fields was chosen to maximise overlap with the radio detections.}
\label{fig:1}
\end{figure*}

\section{Data and measurements}
\label{sample}

We describe our sample selection and characterization of some important aspects of the structures
(Table~\ref{table1}).
A complete description of the fields will be given elsewhere (E. Daddi et al. in preparation).

\subsection{Sample selection}

The 9 structures include several already well known: Cl-1449 (e.g., Valentino et al. 2016), Cl~1001 (Wang et al. 2016), RO-1001 (Daddi et al 2021; D21 hereafter),   CC-0958 (Strazzullo et al. 2015), and XLSSC122 (Mantz et al.  2018). We present here two new radio  overdensities (RO-0959 and RO-0958), selected following Daddi et al. (2017). We also include two \lya\ blobs, SXDS-N-LAB1 (Matsuda et al. 2011; Subaru narrow-band imaging), and FVX-LAB  (from our own narrow-band imaging in COSMOS). 

\begin{table*}
\centering
\caption{Galaxy groups and clusters used in this work. 
\label{table1}}
\begin{tabular}{lrrrrrrrc@{\hspace*{1mm}}c@{\hspace*{1mm}}c@{\hspace*{1mm}}c}
ID & RA & DEC & $z$  & log(M$_{\rm DM}$) &  log($\rm L_{\rm Ly\alpha}$) & log(SFR) & log($\rm L_{\rm AGN}$) & log(BAR)  & log($\frac{M_{\rm stream}}{M_{\rm DM}}$)  & T$_{\rm int}$ & SB corr \\
 & & &  & (M$_{\odot}$) & ($\rm ergs\,s^{-1}$) & (M$_{\odot}$ yr$^{-1}$) & ($\rm ergs\,s^{-1}$) & (M$_{\odot}$ yr$^{-1}$)  &  & h  & dex\\ 
 & & &(1)  & (2) & (3) & & (4) & 	\\ \hline
XLSSC122 & 02:17:44.19 &  -03:45:31.5& 1.98 &     14.2 & $<43.0$ & $<2.3$ &   $<45.3$ & 4.4 & -1.8 & 0.75 & --\\
Cl-1449 & 14:49:14.05 & 08:56:24.6 & 1.992 & 13.8 & $\;43.5$ & $2.8$  & 45.5 & 3.9 & -1.4 & 3.6 & --\\
FVX-LAB & 09:58:42.32  &  02:00:39.3  & 2.194    &     13.0   &    43.6   &     2.1    &    45.3    &   3.1   &      -0.4 & 1.0 & -- \\
Cl-1001  & 10:00:57.18 & 02:20:08.4 &     2.501    &    13.9   &    43.6   &    3.2     &     45.0$^*$   &    4.2 &        -0.9 & 5.0 & --\\
CC-0958  & 09:58:52.97  & 01:58:02.8     &     2.515    &    13.6   &    43.9   &    2.3   &   $<44.6^*$   &    3.9 &         -0.6 & 2.0 & --\\
RO-1001 & 10:01:23.06  & 02:20:04.9  &     2.915    &    13.6   &    44.1   &    3.1   &    44.9$^*$     &      4.0   &      -0.1 & 8.5 & --\\
RO-0959   & 09:59:59.48 &  02:34:41.7 &     3.096   &     12.8   &     44.0   &     3.2   &    45.1   &    3.1   &        0.9 & 1.5 & $0.07$\\
SXDS-N-LAB1   & 02:18:21.31 & -04:42:33.1 &    3.109    &    13.1    &    44.0   &    2.2   &   $<44.9^*$    &   3.4   &        0.6 & 1.0 & $0.08$\\
RO-0958   & 09:58:19.79 & 02:36:10.1 &     3.295   &     12.9   &    43.3    &   3.2   &    45.5$^*$   &    3.2   &        1.0 & 1.25 & $0.18$ \\
\hline
\end{tabular}
\\ Notes: (1) the redshift is from the luminosity weighted \lya\ emission for all but XLSSC122 where it is from optical spectroscopy (Willis et al. 2020); 
(2) we use $M_{200}$; for XLSSC122 we converted $M_{500}$ into $M_{200}$ with a $\times1.7$ scaling; 
(3) SB corrections are already applied; (4) $^*$ indicate values inferred from \lya\ point-source components (or lack there-of);
\vspace{0.5truecm}
\end{table*}

\subsection{KCWI observations and redshift identification}

All the targets were observed with KCWI during observing runs in January 2018 and February 2019. The data reduction and analysis, including diffuse \lya\ identification and characterisation via adaptive smoothing over a 3$\sigma$ threshold  follows D21's work for RO-1001. We detect giant \lya\ nebulae in all  structures ($>100$~kpc; Figure~\ref{fig:1}), except the most massive/evolved XLSSC122 where we determine a conservative $3\sigma$ upper limit over 1000~km~s$^{-1}$ and 200 arcsec$^2$. We confirm  known nebulae: the \lya\ luminosity of Cl-1449 agrees with Valentino et al. (2016), while SXDS-N-LAB1 is $\times2.4$ brighter than in Matsuda et al. (2011), consistent with its emission being partly redshifted out of their narrow-band filter.

Fig.~\ref{fig:1} shows three-band color-images of 8 structures newly observed in \lya\ (see D21 for RO-1001). This includes  a new giant \lya\ halo discovery inside Cl-1001. 
The photometric redshift of RO-0958 and RO-0959 were confirmed by \lya\ detections implying $z=3.29$~and~3.09, respectively. These have been subsequently confirmed with ALMA from the detection of multiple CO lines (E. Daddi et al. in preparation). CC-0958 had  tentative  $z\sim2.18$ (Strazzullo et al. 2015) but KCWI revealed its giant \lya\ halo at $z=2.51$, still consistent with its photometric redshift.

The \lya\ luminosities in Table~1 are integrated above a redshift dependent surface brightness (SB) of $2\times10^{-18}\times[(1+2.78)/(1+z)]^4$~erg~s$^{-1}$~cm$^{-2}$~arcsec$^{-2}$ to account for SB dimming. We need (small) positive luminosity corrections only for the three $z>3$ structures (Table~1), where observed SB limits are shallower than this threshold. Corrections were estimated using the other 5 profiles as a guide, with uncertainties $<0.05$~dex.

The \lya\ luminosities in Table~1 refer to the extended, diffuse emission only:
for all structures we identified \lya\ components arising from galaxies (generally AGNs, see below), and removed their contribution modelling them with the PSF as all remain unresolved at our resolution (typically 0.6--0.8$''$). This correction is  $\sim15$\% for FVX-LAB and RO-0958, and much smaller elsewhere.

\subsection{Estimates of host halo masses, SFRs and AGN content}

Host halo mass ($M_{\rm DM}$) estimates were already presented elsewhere for several structures (XLSSC122; Cl-1449; Cl-1001; RO-1001), where derivations based on their stellar mass ($M*$) content were confirmed via X-ray luminosities and in two cases from the Sunyaev-Zel'dovich effect (SZ; XLSSC122 in Mantz et al. 2018; Cl-1449 in Gobat et al. 2019). For the other structures we derive estimates from the $M*$  following D21. We consider spectroscopic  members, and those with consistent photometric redshifts (Laigle et al. 2016; Muzzin et al. 2013; Mehta et al. 2018),  spatially coincident within the area of the structure, as gauged by the \lya\ halo  and self-consistently with the implied virial radius. 
The integrated $M*$ above the completeness limit for the redshift is corrected to total using the mass functions in Muzzin et al. (2013), and converted to $M_{\rm DM}$  using van der Burg et al. (2014). For CC-0958 we find a marginal 2.4$\sigma$ detection in Chandra, fully consistent with the $M*$-based estimate. 
For the higher mass systems with $M_{\rm DM}\gtrsim 4\times10^{13} M_\odot$, we estimate $M_{\rm DM}$ uncertainties at the level of 0.2~dex or better, at least in relative terms, by comparing the $M*$, X-ray and SZ derivations.

For the lower-mass systems ($M_{\rm DM}\lesssim 1\times10^{13} M_\odot$)  we check that consistent estimates are derived using the brightest group galaxy, applying the relations in van der Burg et al. (2014) and Behroozi et al. (2013).   
We expect these estimates to be uncertain at the 0.3-0.4~dex level (e.g., Looser et al. 2021). 

Integrated bolometric IR luminosities were presented elsewhere for Cl-1449, Cl-1001 and RO-1001 (refs in Sect.~2.1).  
We derived them for the other structures  using Herschel PACS and SPIRE (plus other submm) observations, as described in D21 for RO-1001. Their uncertainties are small,  below 0.2~dex.  We set a conservative 3$\sigma$ upper limit from non-detections in XLSSC122.  We convert IR luminosities into SFRs following Daddi et al. (2007).

We used ancillary Chandra X-ray catalogs (Civano et al. 2016; Marchesi et al. 2016; Mantz et al. 2018) and SED decomposition (e.g., Jin et al. 2018) to search for known AGNs within the expected virial radius of the structures. AGNs are found for  Cl-1449, FVX-LAB,  and RO-0959. Their bolometric luminosities ($L_{\rm bol, AGN}$) were calculated from the mid-IR torus emission   from SED-fitting (e.g. Jin et al. 2018), or its upper limit. If available and more constraining, intrinsic X-ray luminosities scaled to bolometric (Lusso et al. 2012) were used instead. For these AGNs we find coincident point-like \lya\ emission with a ratio of $\mbox{log}\ L_{\rm AGN, bol}/L_{\rm AGN, Ly\alpha}\sim2.9\pm0.2$,  consistent with Sloan QSO for their narrow \lya\ component  (Vanden Berk et al. 2001; Norman et al. 2002). We use this relation to estimate $L_{\rm AGN, bol}$ from \lya\ point-like galaxy components in RO-1001, Cl-1001 and RO-0958, and upper limits for the reminders. 
For RO-1001 this implies $L_{\rm AGN, bol} \sim 10^{44.9}$~erg~s$^{-1}$ (in Galaxy-C), consistent with D21. 

{ The average AGN/SFR ratio in our structures is consistent with the cosmic average (Mullaney et al. 2012; Delvecchio et al. 2019), excluding strong relative enhancements of either quantity.}

\begin{figure*}[ht]
\centering
\includegraphics[width=8cm,angle=0]{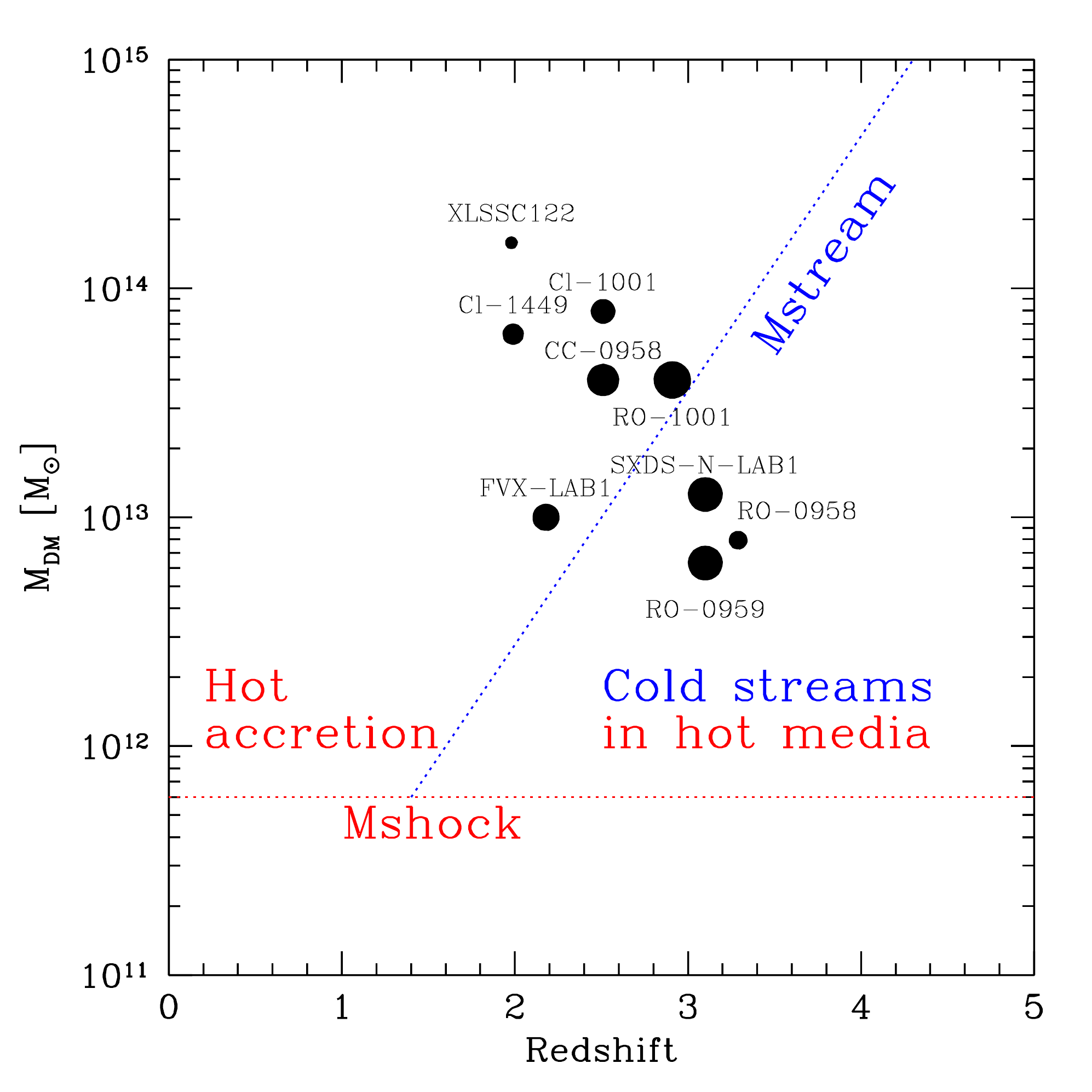}
\includegraphics[width=8cm,angle=0]{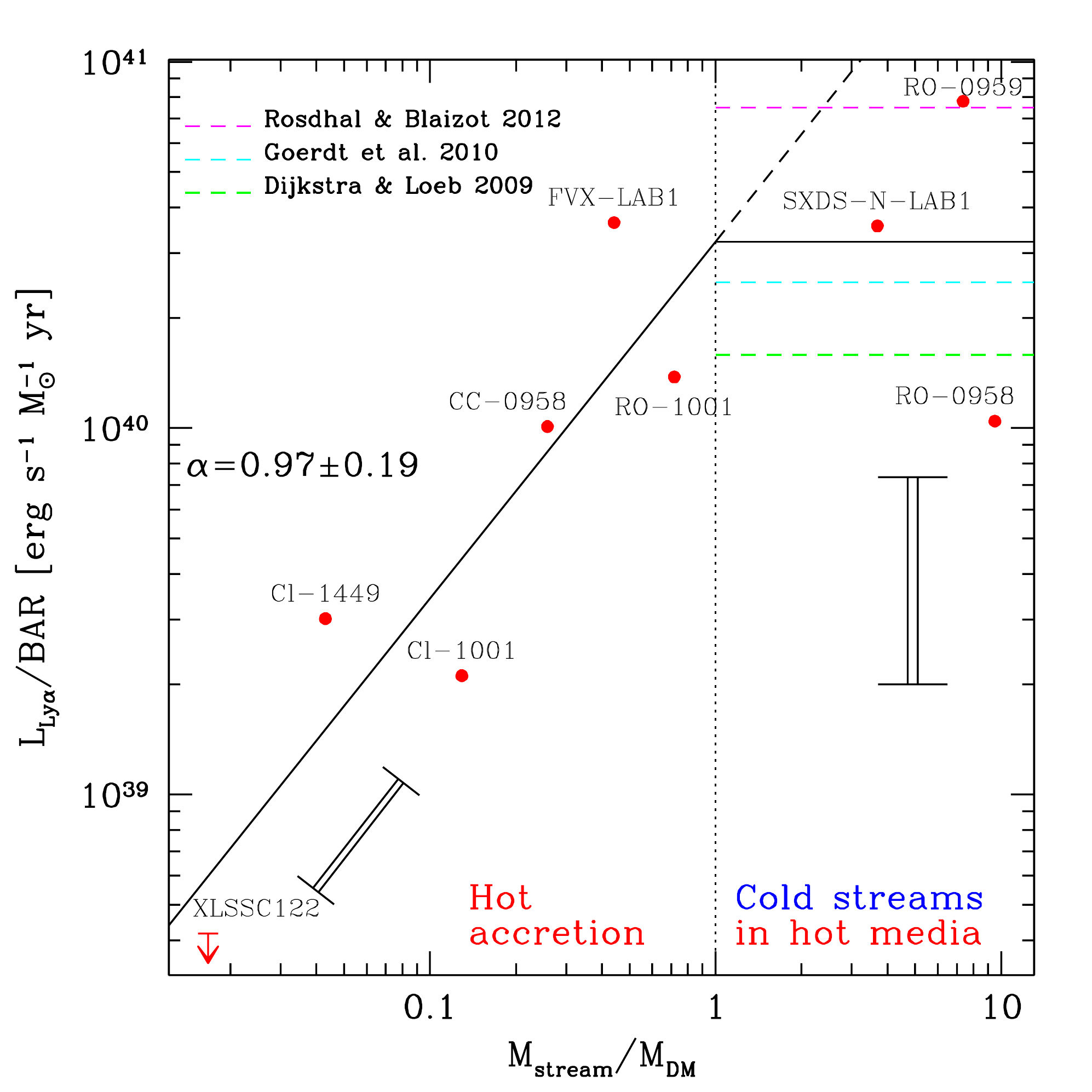}
\caption{(Left:) Our sample in the DB06 diagram. Symbol sizes are proportional to $L_{\rm Ly\alpha}$ (Tab.~1). The blue diagonal line defines $M_{\rm stream}$ (Eq.~\ref{eq:mstream}). Right: the ratio of extended \lya\ luminosity in the structures is plotted versus the $M_{\rm stream}$  to halo-mass ratio. The relation in Eq.~\ref{eq:alpha} is fitted (solid black line). Typical  uncertainties are shown:  0.2~dex along the slope above $M_{\rm stream}$, 0.3~dex along the y-axis below $M_{\rm stream}$. Predictions for $M_{\rm DM}<M_{\rm stream}$  (cold-stream regime) are shown  (colored dashed lines). 
}
\label{fig:2}
\end{figure*}

\begin{figure*}[ht]
\centering
\includegraphics[width=8cm,angle=0]{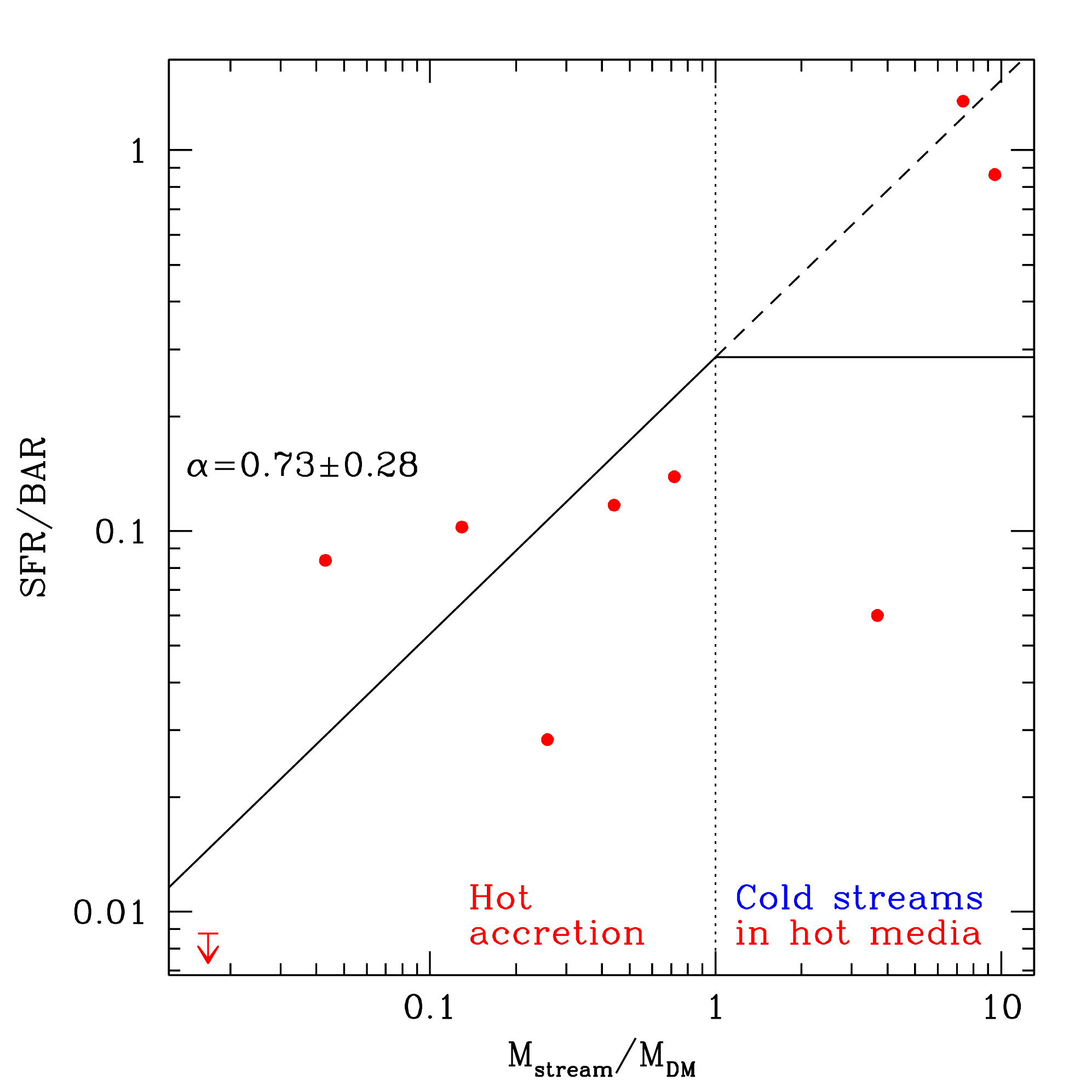}
\includegraphics[width=8cm,angle=0]{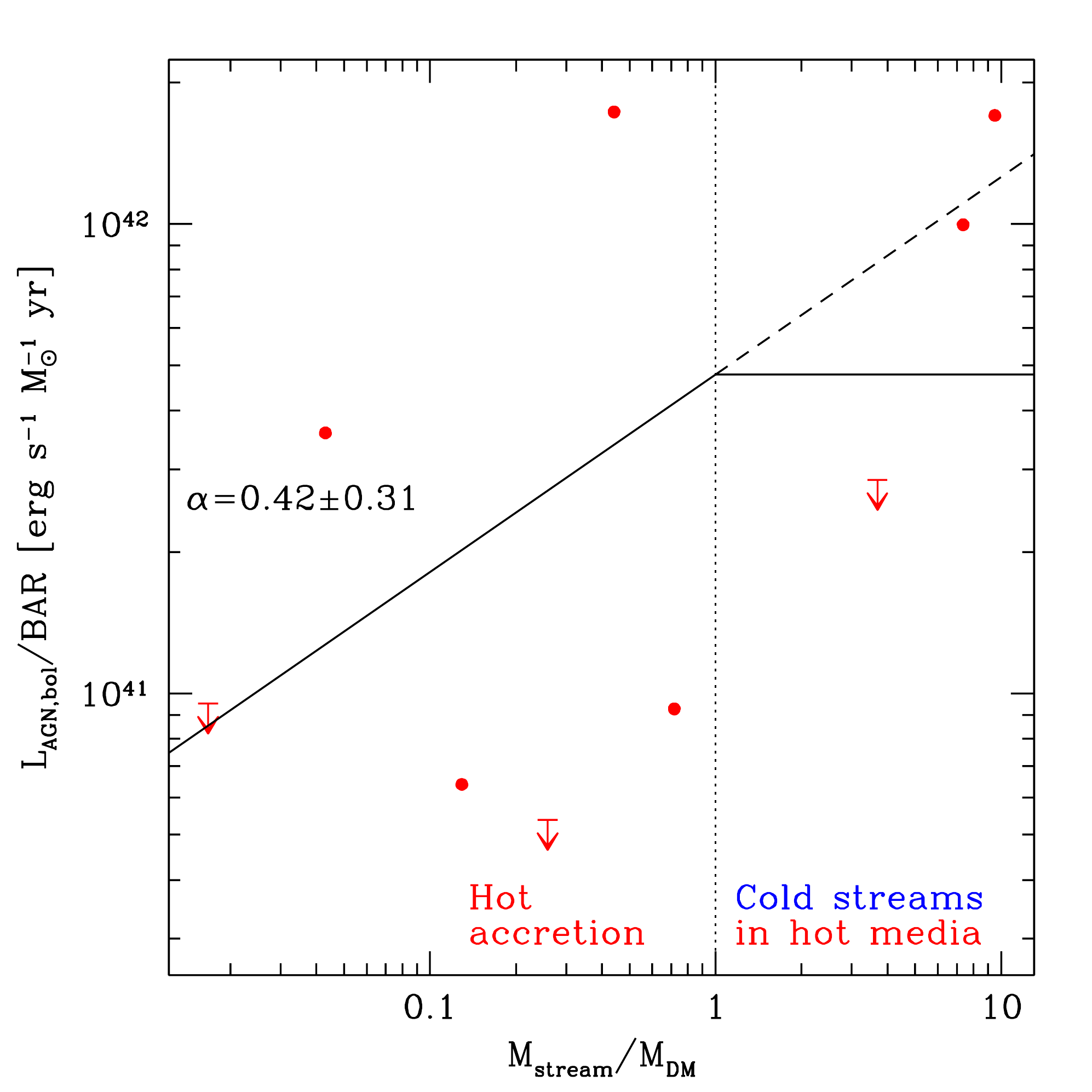}
\caption{
As in Fig.~\ref{fig:2}-right but for SFR (left) and AGN (Right).   Notice the especially poor correlation with $L_{\rm AGN}$/BAR.
}
\label{fig:2b}
\end{figure*}

\section{Results}

We calculate the total halo BAR using Eq.5 from Goerdt et al. 2010 { (using the equivalent formulations from Genel et al. 2008 or Dekel et al. 2013 would not affect our results)}: 

\beq
BAR \simeq 137 (\frac{M_{\rm DM}}{10^{12}M_\odot})^{1.15} (\frac{(1+z)}{(1+3)})^{2.25} M_\odot yr^{-1}
\label{eq:accr}
\eeq

\noindent
and from Fig.7 in DB06 we use $M_{\rm shock}=6\times10^{11}M_\odot$ and  approximate $M_{\rm stream}$  as:

\beq
{\rm log}\ M_{\rm stream} \simeq {\rm log}\ M_{\rm shock} +1.11\times(z-1.4)
\label{eq:mstream}
\eeq

\noindent
{ Fig.~\ref{fig:2}-left shows the location of our sample in the DB06 diagram, spreading across the $M_{\rm stream}$ boundary.
 Fig.\ref{fig:2}-right shows the ratio of $L_{\rm Ly\alpha}$ to BAR for our sample as a function of $M_{\rm stream}/M_{\rm DM}$ ratio.  }

According to theory, cold streams should feed galaxies and halos for $M_{\rm DM} < M_{\rm stream}$ at $z>1.4$, where cold accretion  is equal to total accretion as in Eq.~\ref{eq:accr}, while for $M_{\rm DM} > M_{\rm stream}$ one can expect a smooth transition where an increasingly lower fraction of accretion will be cold (DB06). { We define: 

\beq
BAR_{\rm cold} \simeq
\left\{
	\begin{array}{ll}
	BAR (\frac{M_{\rm stream}}{M_{\rm DM}})^{\alpha}	  &  M_{\rm DM} > M_{\rm stream} \\
		BAR & M_{\rm DM} \lesssim M_{\rm stream}
	\end{array}
\right.
\label{eq:xbar}
\eeq

We  model observables that are expected to be dependent on the availability of cold fuel as:

\beq
L_{Ly\alpha} = C_{Ly\alpha}  \times BAR_{\rm cold} 
\label{eq:alpha}
\eeq

\noindent
where $C_{Ly\alpha}$ is a constant { and the slope from Eq.~3 becomes $\alpha_{\rm Ly\alpha}$.} Similar relations are considered for integrated SFRs and $L_{\rm bol,\ AGN}$, with their modulation slopes $\alpha_{\rm SFR}$, $\alpha_{\rm AGN}$, and constants $C_{\rm SFR}$ and $C_{\rm AGN}$.

Fig.~\ref{fig:2}-right shows a behavior quite consistent with Eq.\ref{eq:alpha}. { A linear fit to the data attempting to constrain its two free parameters returns $\alpha_{Ly\alpha}=0.97\pm0.19$ and ${\rm log}\  C_{Ly\alpha}=40.51\pm0.16$ with a scatter of  0.30~dex. Paired bootstrap (with replacements) implies similar uncertainties. The modulation of decreasing \lya\ luminosity to accretion ratio $\alpha_{Ly\alpha}$, when halo mass is larger than $M_{\rm stream}$, is hence detected at  5$\sigma$.}
{ $C_{Ly\alpha}$  is consistent within a factor of two of model predictions  (Dijkstra \& Loeb 2009; Goerdt et al. 2010; Rosdhal \& Blaizot 2012; Fig.~\ref{fig:2}-right), where roughly 1\% of the gravitational energy goes into \lya. }
{ Note that we are {\it assuming} the flattening in the right side of Fig.~\ref{fig:2}-right, not measuring it. However, if we were to extrapolate the linear fit in Fig.~\ref{fig:2}-right above $M_{\rm stream}/M_{\rm DM}>1$ (dashed line), the observed $L_{\rm Ly\alpha}$ there would deviate by 4.8$\sigma$ in one case (RO-0958) and 1.5$\sigma$ in the other two, all the three weaker than predicted by the extrapolation. This is  unlikely to happen by chance,  supporting the assumed flattening.}

For the ratio SFR/BAR we find a consistent behavior (Fig.~\ref{fig:2b}-left), but less significant (2.6$\sigma$): $\alpha_{SFR}=0.78\pm0.28$ and ${\rm log}\  C_{SFR}=-0.54\pm0.23$, with a 0.45~dex scatter. { Below  $M_{\rm stream}$ (cold-stream regime)  some 20--50\% of the cold accretion goes into SFR, on average (with a scatter of $\times3$). These fractions appear reasonable (Dekel et al. 2009), given that some reduced efficiency seems inevitable as not all the cold gas will be rapidly consumed. 
Above $M_{\rm stream}$ (hot regime),
SFR/BAR in $z\sim2$--2.5 structures is higher by $\times3$--10 than predicted by Behroozi et al. (2013).} 

For AGNs we find only a 1.3$\sigma$ hint of a trend (Fig.~\ref{fig:2b}-right), still consistent with a slope $\sim1$ within 2$\sigma$: $\alpha_{AGN}=0.42\pm0.31$ and ${\rm log}\ C_{AGN}=41.70\pm0.26$, with rms of 0.51~dex.

\section{Discussion}

\subsection{Reliability of the detection}

From our admittedly small and inhomogeneous sample we find  observational evidence in support of the cold-stream to hot-accretion transition  predicted by theory. 
It is crucial to assess its validity. 
Fig.~\ref{fig:2} contains relations among various quantities including estimates based on observables ($L_{\rm Ly\alpha}$ and $M_{\rm DM}$) and  calculated from theory ($M_{\rm stream}$ and BAR; Eqs.~\ref{eq:accr}~and~\ref{eq:mstream}) that depend in turn on $z$ and $M_{\rm DM}$. Redshift   and  \lya\ luminosity  errors are negligible with respect to the scatter observed in the fit. Hence, $M_{\rm DM}$ is the most critical quantity in our analysis. 
For $M_{\rm DM} < M_{\rm stream}$ (cold-stream regime), $L_{\rm Ly\alpha}/BAR$ is set to constant ($C_{Ly\alpha}$; Eq.~\ref{eq:accr}), because cold BAR equals total BAR. 
The uncertainty in $C_{Ly\alpha}$ is 0.16~dex, 
consistent with  $M_{\rm DM}$ error propagation (0.3/$\sqrt5$~dex), indirectly confirming the error estimates. 
In the hot regime ($M_{\rm DM} > M_{\rm stream}$), the three structures with highest $M_{\rm DM}/M_{\rm stream}$ are the ones with the best halo mass determination, estimated from X-ray and SZ measurements. By themselves, they set a slope $\alpha_{\rm Ly\alpha}\approx1$ respect to the rest of the sample. 
This  simplifies to first order Eqs.~~\ref{eq:accr}~and~\ref{eq:mstream} to $L_{\rm Ly\alpha}\propto M_{\rm stream}$, and we find indeed $L_{\rm Ly\alpha}/M_{\rm stream} = 10^{30.75\pm0.20}$~erg~s$^{-1}$~$M_\odot^{-1}$ with just 0.20~dex scatter, implying that $L_{\rm Ly\alpha}$ grows by 1~dex over $\Delta z=1$, similarly to $M_{\rm stream}$. Hence in the hot-regime (and for $\alpha\approx1$) a precise determination of $M_{\rm DM}$ is not required to measure $\alpha$. 
Our conclusions are thus not critically dependent on  $M_{\rm DM}$ uncertainties. 

\begin{figure*}[ht]
\centering
\includegraphics[width=8cm,angle=0]{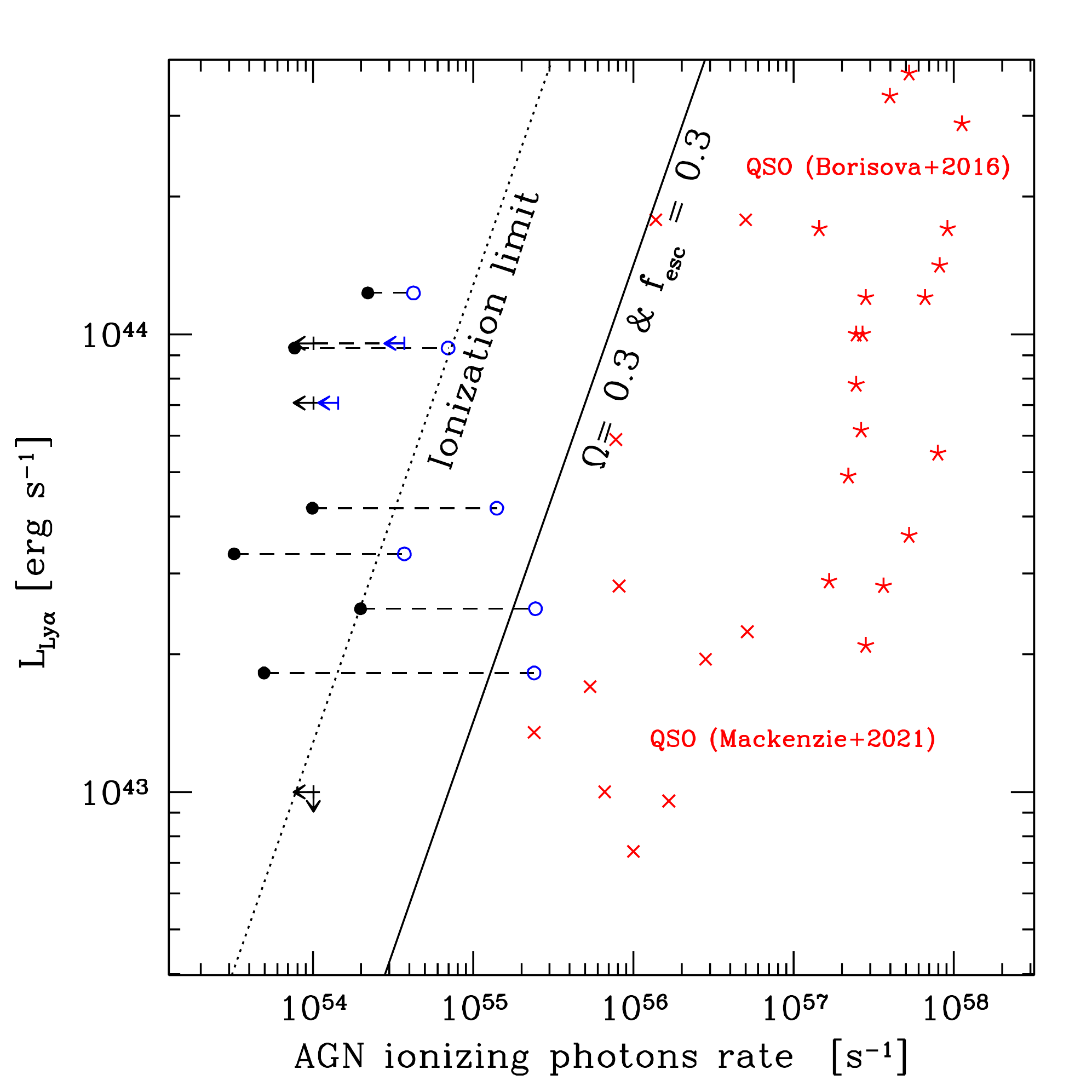}
\includegraphics[width=8cm,angle=0]{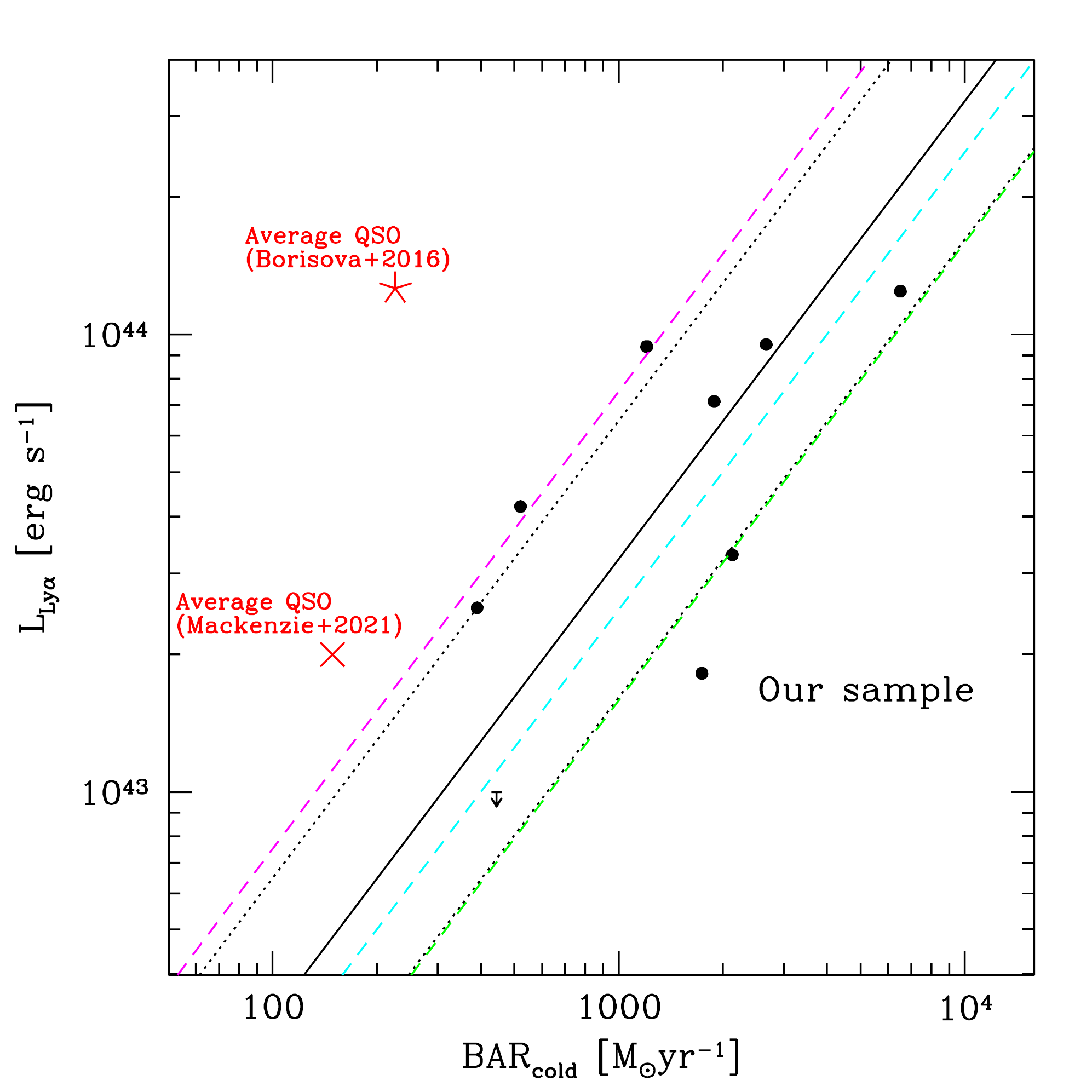}
\caption{(Left) The \lya\ luminosity versus  the AGN  ionizing photon rates for our sample (black points estimated from the  ultraviolet luminosity, blue empty point connected by dashed lines are computed from $L_{\rm bol, AGN}$, see text).  The diagonal lines show the  \lya\ luminosity that AGN can ionise: theorethical maximum (dotted) and (solid) assuming a 30\% escape fraction  (Smith et al. 2020) and an opening angle $\Omega=30$\% (Simpson et al. 2005). (Right) The \lya\ luminosity versus the cold accretion rate, as resulting from Eq.~\ref{eq:xbar}. 
The solid (dotted) line(s) show the average linear trend  (1$\sigma$ range). The colored dashed lines are models as in Fig.~\ref{fig:2}-right (Dijkstra \& Loeb 2009; Goerdt et al. 2010; Rosdhal \& Blaizot 2012).
{ In both panels QSO-selected \lya\ nebulae are shown, individually in the left-panel and averaged in the right-panel where the QSOs' average hosting $M_{\rm DM}$, hence BAR, are estimated from Eftekharzadeh et al. (2015). }
}
\label{fig:3}
\vspace{1.1truecm}
\end{figure*}

\subsection{Physical interpretation}

{ It is important to question whether the correlation shown in Fig.~\ref{fig:2}-right arises from a direct link between  cold accretion rate and \lya\ luminosity (Fig.~\ref{fig:3}-right), or an indirect one, with accretion regulating SFR and $L_{\rm bol, AGN}$ that in turn determine \lya\ emission, e.g. by photo-ionization and subsequent recombinations (see D21 for more discussions). 
If the latter,  it would be difficult to explain how \lya\ can define a  tighter relation to accretion than SFR and AGN if \lya\ was a byproduct of these  quantities. 
D21 already suggested cold accretion as the main source of \lya\ powering in RO-1001, rather than AGN or SFR. 
However, it is worth reconsidering the matter here, with a larger sample of structures. 

Starting from SFRs, 
 basically in all cases, the most highly star-forming members  are heavily dust-extinguished (Fig.~\ref{fig:1}); their  contribution to \lya\ photo-ionization from UV-unattenuated SFR appears negligible, as in D21. 

AGN photoionization should be more carefully considered as a plausible source for powering \lya, at least in some of our structures. We estimate \lya\ photo-ionisation rates from $L_{\rm bol, AGN}$ (measurements or upper limits) using  bolometric corrections to ultraviolet (Trakhtenbrot \& Netzer 2012) and the Type1 QSO average spectrum (Lusso et al. 2015). 
 Fig.~\ref{fig:3}-left shows that the maximum theorethical AGN ionizing radiation  is  potentially sufficient to power $L_{\rm Ly\alpha}$ only in 4 of the 9 structures. 
 When considering that at these luminosities the Type~1 opening angle is expected $\sim30$\% (Simpson et al. 2005; possibly too large for our sample where no unobscured AGN is found from the 9 structures, Fig.~\ref{fig:3}), and the Lyman Continuum escape  $\sim30$\% (Smith et al. 2020; also likely an overestimate given that our AGNs are embedded in high dust optical dephts based on ALMA detections), only two structures remain marginally viable to be fully AGN photo-ionized, neglecting further geometrical effects (Valentino et al. 2016).

 Our sample is compared in Fig.~\ref{fig:3} to \lya\ nebulae selected around QSOs, from Borisova et al. (2016) and Mackenzie et al. (2021). Their \lya\ photo-ionization rates to $L_{\rm Ly\alpha}$ ratios are 1--2 orders of magnitude larger, on average, with respect to structures in our sample (Fig.~\ref{fig:3}-left). 
Also, 
  QSOs are significantly overluminous in their average ratio of \lya\ emission  to cold accretion rates (Fig.~\ref{fig:3}-right). It is thus possible that our structures' \lya\ powering is not coming from photo-ionization, as for the QSOs.}
  
  Although these figures show scatter, and the impact of AGNs (and SFR) might vary and be somewhat larger in individual cases, the favored scenario is currently that their contribution is not dominant to \lya\ in massive groups and clusters at $2<z<3.3$, at least for those in the cold-stream regime. 

In the hot-regime, if $\alpha_{SFR}$ and $\alpha_{AGN}$ were to be truly flatter than $\alpha_{Ly\alpha}$, this could suggest the contributions from SFR and AGN to \lya\ being increasingly important with growing $M_{\rm DM}/M_{\rm stream}$. Such a behaviour would be physically motivated by the longer timescales required for SFR quenching and residual gas consumption (of order 100-300~Myr; AGNs just reflecting the SFR with increased stochasticity), with respect to \lya\ from accretion that is likely a more instantaneous measure. Determining $\alpha_{SFR}$  ($\alpha_{AGN}$) to sufficient precision, e.g. $5\sigma$, would require larger samples of $\sim 50$ ($\sim 200$) massive groups and clusters. 

\subsection{Predicting \lya\ luminosities for dark matter halos}

{The good fit of Eqs.~3-4 to Fig.~\ref{fig:2}-right is encouraging in terms of using \lya\  to trace accretion, hence ultimately halo masses, e.g. for unveiling  dark matter halo locations in  proto-clusters. 
Using the measured $C_{\rm Ly\alpha}$, Eq.~\ref{eq:accr}-\ref{eq:alpha} can be re-written in the cold-stream regime as:}

\beq
\mbox{log}\ L_{Ly\alpha}/\mbox{erg s}^{-1} \sim 43.6+ \mbox{log}\ \frac{M_{\rm DM}}{10^{13}M_\odot} + 2.25\ \mbox{log}\frac{1+z}{1+3},
\eeq

\noindent
depending (quasi)linearly on $M_{\rm DM}$. $L_{Ly\alpha}$ can increase only until $M_{\rm DM}$ reaches $M_{\rm stream}$. 
From that point on,  in the hot-accretion regime, $L_{\rm Ly\alpha}$ is roughly constant depending only on redshift, regardless of how large $M_{\rm DM}$ can reach (because $\alpha_{\rm Ly\alpha}\sim1$). 
The data (although using only 6 objects) allow us to fix the numerical parameters as: 

\beq
 L_{Ly\alpha}/\mbox{erg s}^{-1} \sim 10^{42.6} (\frac{1+z}{1+1.4})^{\sim7} \ \  \mbox{for}\ M_{\rm DM}>M_{\rm stream}
\eeq
\smallskip

\noindent
implying that the typical \lya\ nebula  in the hot-regime (and saturation level for the cold-regime) is $\approx10^{43.3}$~erg~s$^{-1}$ at $z=2$ and $\approx10^{44}$~erg~s$^{-1}$ at $z=3$. 
The exponent recovered is lower than what fixed by theory (Eqs.~1~and~2), perhaps suggesting a less steep $M_{\rm stream}$ redshift dependence (see discussion in Dekel et al. 2009 appendixes). { These predictions can be tested with larger samples. }
\smallskip

In conclusion, we report widespread giant \lya\ nebulae in massive groups/clusters at $2<z<3.3$, the only non-detection being the most evolved cluster. The \lya\ luminosity is  a smoothly decreasing fraction of the total baryonic accretion onto these massive halos, for the range where models predict that cold-streams should progressively cease feeding halos, thus supporting these models. 

\begin{acknowledgements}
We thank Dawn Erb for sharing calibrations,  Sebastiano Cantalupo for  his CubEx code and discussions, and  the referee for a constructive report. RMR acknowledges GO-15910.002 from the Space Telescope Science Institute. Data were obtained at the W. M. Keck Observatory,  operated as a scientific partnership among the California Institute of Technology, the University of California and the National Aeronautics and Space Administration, made possible by the generous financial support of the W. M. Keck Foundation. The authors also acknowledge the indigenous Hawaiian community and are grateful for the opportunity to collect data from the summit of Maunakea.
\end{acknowledgements}

\end{document}